\documentclass[12pt]{article}
\usepackage{euscript,amsmath,amssymb,latexsym,amsthm,cite}

\newcommand{\be}{\begin{equation}}
\newcommand{\ee}{\end{equation}}
\newcommand{\bea}{\begin{eqnarray}}
\newcommand{\eea}{\end{eqnarray}}

\theoremstyle{definition}
\newtheorem*{remark}{Remark}

\theoremstyle{theorem}
\newtheorem*{proposition}{Proposition}

\begin{document}
\title{
\vspace{1cm} {\bf Functional Classical Mechanics and \\
Rational Numbers}
 }
\author{A.\,S.~Trushechkin and I.\,V.~Volovich
 \\
{\it  Steklov Mathematical Institute}
\\ {\it Gubkin St.8, 119991 Moscow, Russia}
\\ email:\:\texttt{trushechkin@mi.ras.ru,\, volovich@mi.ras.ru}}

\date {}
\maketitle

\begin{abstract}
The notion of microscopic state of the system at a given moment of
time as a point in the phase space as well as a notion of trajectory
is widely used in classical mechanics. However, it does not have an
immediate physical meaning, since arbitrary real numbers are
unobservable. This notion leads to the known paradoxes, such as the
irreversibility problem. A ``functional" formulation of classical
mechanics is suggested. The physical meaning is attached in this
formulation not to an individual trajectory but only to a ``beam" of
trajectories, or the distribution function on phase space. The
fundamental equation of the microscopic dynamics in the functional
approach is not the Newton equation but the Liouville equation for
the distribution function of the single particle.  The Newton
equation in this approach appears as an approximate equation
describing the dynamics of the average values and there are
corrections to the Newton trajectories.  We give a construction of
probability density function starting from the directly observable
quantities, i.e., the results of measurements, which are rational
numbers.

\end{abstract}

\section{Introduction}

The conventional widely used concept of the microscopic state of the
system in classical Newtonian mechanics \cite{Arn} at some  moment
of time as the point in phase space, as well as the notion of
trajectory and the microscopic equations of motion have no direct
physical meaning, since arbitrary real numbers are unobservable.
Observable physical quantities are only presented by rational
numbers \cite{ Vol1, Vol2, Zel}, see also the discussion of concepts
of space and time in \cite{DKKV,Khr1,VVZ,Zel,Vol2,Vol1,Var}.

In \cite{Vol3} it was suggested a ``functional" formulation of
classical mechanics. The fundamental equation of the microscopic
dynamics in the functional approach is not the Newton equation, but
the Liouville equation for the distribution function of a single
particle.  The Newton equation in this approach appears as an
approximate equation describing the dynamics of the average values,
and there are corrections to the Newton trajectories. The functional
formulation of classical mechanics gives also an approach to the
solution of the irreversibility problem.

In this note we give a construction of the probability density
function starting from the directly observable quantities, i.e., the
results of measurements, which are rational numbers.

\section{States and observables in  \\ functional classical mechanics}

Usually in classical mechanics the motion of a point body is
described by a trajectory in the phase space, i.e., the values of
the position and momentum as functions of time, which are solutions
of the equations of Newton or Hamilton.

However,  this mathematical model is an idealization of the physical
process, rather far separated from reality. Every physical body has
the spatial dimensions, such a mathematical point gives only an
approximate description of the physical body. The mathematical
notion of a trajectory does not have a direct physical meaning,
since it uses  arbitrary real numbers, i.e., infinite decimal
expansions, while the observation is only possible, in the best
case, in rational numbers, and even among them only with some error.
Therefore, in the  ``functional" approach to classical mechanics, we
are not starting from the Newton equation, but with the Liouville
equation.

Consider the motion of a classical particle along a straight line in
the potential field. The general case of many particles in the
3-dimensional space is discussed below. Let $(q, p)$ be coordinates
on the plane $\mathbb{R}^2$  (phase space), $t\in\mathbb{R}$ is
time. The state of a classical particle at time $t$ will be
described by a function $\rho=\rho (q, p, t)$, it is the density of
the probability that the particle at time $t$ has the position $q$
and momentum $p$.

The description of a mechanical system with the help of probability
distribution function $\rho=\rho (q, p, t)$ does not necessarily
mean that we are dealing with a set of identically prepared ensemble
of particles. Usually in probability theory one considers  an
ensemble of events  and a sample space. But we can use the
description with the function $\rho=\rho (q, p, t)$ also for
individual bodies, such as planets in astronomy (the phase space in
this case the 6-dimensional). In this case one can think on the
``ensemble" of different astronomers which observe the planet, or on
the  ``ensemble" of different models of behaviour of a given object
for one ``intelligent" observer. Actually, it is implicitly always
dealt with the function $\rho=\rho (q, p, t)$ which takes into
account the inherent uncertainty in the position and momentum of the
body.

The specific type of function $\rho$ depends on the method of the
preparation of the state of the classical particle at the initial
time and the type of potential field. When $\rho=\rho (q, p, t)$ has
sharp peaks at $q = q_0$ and $p = p_0$, we say that the particle has
the approximate values of the position and momentum $q_0$ and $p_0$.

Emphasize that  the exact determination of the position and momentum
can not be done not only in quantum mechanics, where there is the
Heisenberg uncertainty relation, but also in classical mechanics.
Always there are some errors in setting the position and momentum.
There are classical uncertainty relations:
$$\Delta q\Delta p>0,$$ i.e., the uncertainty (errors of observation)
in the determination of the position and momentum is always positive
(nonzero). The concept of arbitrary real numbers, given by the
infinite decimal series, is a mathematical idealization, such
numbers cannot be measured in the experiment.

Therefore, in the functional approach to classical mechanics the
concept of precise trajectory of a particle is absent, the
fundamental concept is a distribution function $\rho=\rho (q, p,
t)$, and $\delta$-function as a distribution function is not
allowed.

We assume that the continuously differentiable and integrable
function $\rho=\rho (q, p, t)$ satisfies the conditions:
\begin{equation}\label{rho}
\rho \geq 0,~~\int_{\mathbb{R}^2}\rho (q,p,t)dqdp=1,~ t\in
\mathbb{R}\,.
\end{equation}

If $f = f (q, p)$  is a function on the phase space, the average
value of $f$ at time $t$ is given by the integral
\begin{equation}\label{int1}
\overline{f}(t)=\int f(q,p)\rho (q,p,t)dqdp\,.
\end{equation}
In a sense, we are dealing with a random process $\xi(t)$ with
values in the phase space. Motion of a point  body along a straight
line in the potential field will be described by the equation
\begin{equation}\label{Lio}
\frac{\partial \rho}{\partial t}=-\frac{p}{m}\frac{\partial
\rho}{\partial q}+\frac{\partial V(q)}{\partial q} \frac{\partial
\rho}{\partial p}\,.
\end{equation}
Here $V(q)$  is a potential field and $m>0$ is the mass of the body.

Equation (\ref{Lio}) looks like the Liouville equation, which is used
in statistical physics to describe a gas of particles, but here we
use it to describe a single particle.

The characteristics equations  for (\ref {Lio}) are  Hamilton's
equations
\begin{equation}\label{char3}
\dot{q}=\frac{\partial H}{\partial p},\,\,\dot{ p}=-\frac{\partial
H}{\partial q}\,,
\end{equation}
where the Hamiltonian is
\begin{equation}\label{ham3}
H=\frac{p^2}{2m}+V(q)\,.
\end{equation}
Emphasize  again that the Hamilton equations (\ref{char3}) in the
current functional approach to mechanics do not describe directly
the motion of particles, but they are only the characteristics
equations for the Liouville equation (\ref {Lio}).

If the distribution $\rho_0(q,p)$ for $t=0$ is known, we can
consider the Cauchy problem for the equation (\ref{Lio}):
\begin{equation}\label{cau}
\rho |_{t=0}=\rho_0(q,p)\,.
\end{equation}
Consider the case when the initial distribution has the Gaussian
form:
\begin{equation}\label{gauss}
\rho_0 (q,p)=\frac{1}{\pi
ab}e^{-\frac{(q-q_0)^2}{a^2}}e^{-\frac{(p-p_0)^2}{b^2}}\,.
\end{equation}
At sufficiently small values of the parameters $a> 0$ and $b> 0$ the
particle has the position and momentum close to the $q_0$ and $p_0$.
For this distribution the average values of the position and
momentum are
\begin{equation}\label{mean1}
\overline {q}=\int q\rho_0 (q,p)dqdp=q_0\,,~~\overline{p}=\int
p\rho_0 (q,p)dqdp=p_0\,,
\end{equation}
and the dispersion
\begin{equation}\label{disp1}
\Delta q^2=\overline{(q-\overline {q})^2}=\frac{1}{2}a^2,~~\Delta
p^2=\overline{(p-\overline {p})^2}=\frac{1}{2}b^2\,.
\end{equation}
For the free motion ($V=0$) we get
\begin{equation}\label{coor2} \overline {q}(t)=q_0+\frac{p_0}{m}t
\,,~~\overline{p}(t)=p_0\,,
\end{equation}
and the dispersion  increases with time:
\begin{equation}\label{disp2}
\Delta q^2(t)=\frac{1}{2}(a^2+\frac{b^2t^2}{m^2}).
\end{equation}
Even if the particle was arbitrarily well localized ($a^2$ is
arbitrarily small) at $t = 0$, then at sufficiently large times $t$
 the localization of the particle becomes meaningless, there is a
{\it delocalization} of the particle which accounts for
irreversibility.

Corrections to the Newton's trajectories for the nonlinear coupling
are computed in \cite{Vol3}.
\section{Probability density function and rational numbers}

The probability density function is real-valued. However, our
initial point was that the real numbers are unobservable. Does the
use of real-valued probability density function as a fundamental
notion of mechanics contradict to our initial thesis? In this
section we construct the probability density function (formula
(\ref{EqDensMean})) starting from the directly observable quantities
using the methods of mathematical statistics. An important point is
that the probability might be a  real number and this is admissible,
since the probabilities are not directly observable.

The directly observable quantities are the results of measurements
and they are rational numbers. Consider a measurement of an
observable $\mathbb X$. For simplicity we consider a one-dimensional
observable, the generalization is simple. Every measurement device
has an error, which must be taken into account. Roughly speaking,
the measurement errors can be divided into two types: systematic and
random errors \cite{Taylor}. If we perform repeated measurements,
the systematic error does not change and random error changes
randomly (we do not consider the part of the error that changes
regularly, because it can be excluded by the statistical methods).
It is natural to model the random error by a random variable. There
is no theory of systematic error (``In fact, the only theory of
systematic errors is that they must be identified and reduced until
they are much less than the required precision... However, this goal
is often not attainable'' \cite{Taylor}). Therefore, although the
systematic error is constant, our ignorance of the systematic error
is also modeled by a random variable, because of the absence of
another theory.

Thus, the result of a measurement is a random variable $X$,
 it is rational-valued. Moreover, since the precision
 (sensitivity) of every instrument is finite, $X$ takes
 values not on the whole field of rational numbers $\mathbb Q$,
 but rather on the
 lattice, $X\in\frac pq\mathbb Z$, where the rational fraction $\frac pq$ is
 the measuring sensitivity of the instrument and $\mathbb Z$ are integers.
 So, the probabilities
$p_m=\Pr[X=\frac{p}{q}m]$, $m\in\mathbb Z$, are defined. The
probabilities $p_m$ are real and this is admissible, since the
probabilities are not directly observable. They can be considered as
a limit of relative frequencies:
$$p_m=\lim_{n\to\infty}\frac{n_m}n\quad\text{in probability, i.e.,}\quad
\lim_{n\to\infty}\Pr\left[\frac{n_m}n-p_m\right]=0,$$ where $n$ is
the number of experiments and $n_m$ is the number of experiments
where the realization of $X$ is equal to $\frac{p}qm$. This is the
law of large numbers \cite{Gnedenko}. The fact that probabilities
are real numbers, actually, is not surprising, since the limit of
rational sequences is not necessarily a rational number.

Now consider the dynamics. Let us measure the observable $\mathbb X$
once again at some moment of time $t>0$. We want to predict the
probabilities of the results of this measurement on the condition
that we know the result of the measurement at time $t=0$. If we
describe  a state as a sum of delta functions  and solve the
Liouville equation with such initial conditions (this is equivalent
to Newton`s equation), we will get incorrect predictions. For
example, consider the free motion on the real line. Assume that at
time $t=0$ we obtained that momentum is equal to zero with the
precision allowed by our instrument. Then we can conclude that at
any time $t>0$ the particle still will be in its initial position.
But in general this not true, since the momentum can be very small
(smaller than our measuring sensitivity), but not zero. In this
case, if $t$ is large enough, the position of the particle can be
changed considerably.

Thus, in order to take the growth of the error with the time
into account, we must consider the states as continuous distributions.
Let us assign some continuous real-valued random
variable $\widetilde X$ to our discrete random
variable $X$. Let $\widetilde X$ be distributed
according to some probability density
function $\rho_*(x)$ which satisfies the condition
\begin{equation}\label{EqDisContPr}
p_m=\int_{\frac pq(m-\frac12)}^{\frac pq(m+\frac12)}\rho_*(x)dx.
\end{equation}
We assume that $\widetilde X$ is normally distributed:

\begin{equation}\label{EqDensNorm}
\rho_*(x)=\frac{1}{\sqrt{2\pi\sigma^2}}\,e^{-\frac{(x-x^*)^2}{2\sigma^2}},
\end{equation}
where $x^*$ is a mean value (which can, but not necessarily, be
referred as a ``true'' value of the observable $\mathbb X$) and
$\sigma^2$ is a dispersion.
$$\sigma^2=\sigma_{\text{syst}}^2+\sigma_{\text{rand}}^2,$$
where $\sigma_{\text{syst}}^2$ and $\sigma_{\text{rand}}^2$ are the
summands that correspond to the systematic and random error
accordingly. In fact, the further discussion does not depend
critically on the form of distribution. We made an assumption about
normal distribution for simplicity, but there are also some physical
and mathematical reasons to choose this distributions among others.

Again, the notion of real-valued probability density function
$\rho_*$ does not contradict to the thesis that real values are
unobservable, because the probability density function is not an
observable. This is an abstract, theoretical object, which is
useful, because we can approximate the relative frequencies using
the notion of real-valued probability density function:

$$\frac{k\{X\in[a,b]\}}n\approx\int_a^b\rho_*(x)\,dx.$$
Here $k\{X\in[a,b]\}$ is the number of
experiments where the realization of $X$ belongs to $[a,b]$ (for example, $a,b\in\mathbb Q$) and $n$
is the general number of experiments (it is assumed that $n$ is large).

Usually we do not now the expectation value $x^*$ and the dispersion
of random error $\sigma_{\text{rand}}^2$ (and hence, we do not know
the probability density function $\rho_*$, we only assume that it
has the form (\ref{EqDensNorm}) with unknown parameters), but rather
we have to estimate them using the methods of mathematical
statistics. The dispersion of systematic error
$\sigma_{\text{syst}}^2$ is assumed to be known from the measuring
instrument certificate. Let $X^{(1)},\dots,X^{(n)}$ be $n$ copies of
$\widetilde X$, i.e., independent and identically distributed
(according to the probability density function $\rho_*$) random
variables (the results of $n$ measurements). Then the following
formulas are used to estimate the expectation and dispersion of the
random error:

\begin{equation*}
\overline X=\frac{1}{n}\sum_{i=1}^nX^{(i)},\quad
S_\text{rand}^2=\frac{1}{n-1}\sum_{i=1}^n(X^{(i)}-\overline X)^2.
\end{equation*}

It is well-known that random variable $\overline X$ is normally
distributed with the same expectation $x^*$ as every of
$X^{(1)},\dots,X^{(n)}$. The dispersion of the random error in the
estimation of $\overline X$ is reduced by $n$ times and is equal to
$\sigma_{\text{rand}}^2/n$. Hence, its estimation is
$S_{\text{rand}}^2/n$ \cite{Gnedenko}. The dispersion of the
systematic error does not depend on the number of measurements and
still equals to $\sigma_{\text{syst}}^2$ (this is not a rigorous
conclusion, since there is no theory of systematic error and the use
of the formalism of random variables is not very correct for this).
Therefore, the dispersion of the general error of the estimation of
$x^*$ is
\begin{equation*}
S^2=\frac{S_\text{rand}^2}{n}+\sigma_\text{syst}^2.
\end{equation*}
Now we construct the probability density function:
\begin{equation}\label{EqDensMean}
\rho_n(x)=\frac{1}{\sqrt{2\pi S^2}}\,e^{-\frac{(x-\overline
X)^2}{2S^2}}.
\end{equation}
If $n$ is large (in practice, $n>30$ is enough), then
$\rho_n(x)\Delta x$ has the meaning of the probability for the mean
value $x^*$ to belong to the interval $(x-\Delta x,x+\Delta x)$.

\begin{remark}
The last assertion can be understood by physicists, but is not
completely  correct from the mathematical point of view. Since $x^*$
is not a random variable, the probability for it to belong to the
definite interval is whether zero or one. More rigorous formulation
of the assertion is the following: $(1/\sqrt{2\pi})e^{-x^2/2}\Delta
x$ is approximately the probability for $\frac{\overline
X-x^*}{\sqrt{\frac{S^2}{n}}}$ to belong to the interval $(-\Delta
x,\Delta x)$.
\end{remark}

Note that, in fact, $\rho_n$ is a random function, because it
depends on the random variables $\overline X$ and $S^2$.

If $n\to\infty$, then $\overline X\to x^*$, $S_\text{rand}^2/n\to0$, $S^2\to\sigma^2_\text{syst}$ in probability. Denote
$$\rho_\infty(x)=\frac{1}{\sqrt{2\pi
\sigma^2_\text{syst}}}\,e^{-\frac{(x-\overline
X)^2}{\sigma^2_\text{syst}}}.$$ This is also a random function. The
following proposition holds:

\begin{proposition}
$$\lim_{n\to\infty}\Pr\{X^{(n)}\in[a,b]\}=
\int_a^b\rho_\infty(x)dx$$
in probability, i.e.,
$$\lim_{n\to\infty}\Pr\left\lbrace\Pr\{X^{(n)}\in[a,b]\}-
\int_a^b\rho_\infty(x)dx\right\rbrace=0,$$
if $a=\frac pq(m-\frac12)$, $b=\frac pq(l-\frac12)$ for some $m,l\in\mathbb Z$ (in other words, $a,b\in\frac pq\mathbb Z+\frac12$).
\end{proposition}

This is a corollary of the limit theorems of
probability theory
(the law of large numbers and the central limit theorem)
and condition (\ref{EqDisContPr}).

If we perform the repeated measurements, we can predict the
probabilities of the results of the next measurement in the limit
$n\to\infty$ using the constructed probability distribution function
(\ref{EqDensMean}). This justifies the use of the described
construction.

\section{Conclusions}

It is shown that the use of real-valued probability density function
as a fundamental concept of functional classical mechanics does not
contradict to the thesis that the real irrational numbers are
unobservable, since the density function is not a directly
observable value. The construction of the probability density
function based on the rational-valued results of measurements and an
argumentation for this construction are given. It would be
interesting to extend these results to the case of quantum
mechanics, see \cite{VT}.

\section{Acknowledgments}
This work was partially supported by the Russian Foundation for
Basic Research (projects 08-01-00727-a and 09-01-12161-ofi-m),  the
grant of the President of the Russian Federation (project
NSh-3224.2008.1) and by Division of mathematics of RAS.

\end{document}